\input{aipcheck}

\documentclass[
    ,final            
   ,draft            
   ,numberedheadings 
  ]
  {aipproc}

\layoutstyle{6x9}

\newcommand{\km}{${\rm km\,s}^{-1}$}

\newcommand{\hi}{H$\;${\small\rm I}\relax}

\newcommand{\civ}{C$\;${\small\rm IV}\relax}

\newcommand{\nv}{N$\;${\small\rm V}\relax}
\newcommand{\oi}{O$\;${\small\rm I}\relax}

\newcommand{\ovi}{O$\;${\small\rm VI}\relax}

\newcommand{\fuse}{{\it FUSE}}
\newcommand{\degr}{{$^\circ$}}

\begin{document}
\title{The Magellanic System: What have we learnt from \fuse?}


\author{N. Lehner}{
  address={University of Notre Dame, 225 Nieuwland Science Hall
Notre Dame, IN 46556}
}

\begin{abstract}
 I review some of the findings on the Magellanic System 
produced by the {\it Far Ultraviolet Spectroscopic Explorer (FUSE)}\ 
during and after its eight years of service. The Magellanic System with its
high-velocity complexes provides a nearby laboratory that can be used to characterize
phenomena that involve interaction between galaxies, infall and outflow 
of gas and metals in galaxies. These processes are crucial for understanding 
the evolution of galaxies and the intergalactic medium.  Among the \fuse\ successes 
I highlight are the coronal gas about the LMC and SMC, and beyond in the Stream, the 
outflows from these galaxies, the discovery of molecules in the diffuse 
gas of the Stream and the Bridge, an extremely sub-solar and sub-SMC metallicity of the
Bridge, and a high-velocity complex between the Milky Way and the Clouds.
\end{abstract}

\maketitle


\section{Introduction}
The interactions of galaxies and the nearby intergalactic medium (IGM) 
through the accretion of matter onto galaxies or the expulsion
of matter and energy in winds from galaxies are 
crucial for the evolution of both the galaxies and the IGM 
\cite[e.g.,][]{bushouse87,veilleux05,simcoe06}. Interactions between galaxies often produce long tails of 
gaseous matter and galaxy mergers, where starbursts can be triggered \cite[e.g.,][]{larson78}. 
Galactic winds, driven by the energy and momentum deposited into
the interstellar medium (ISM) by massive stars and supernovae or by 
active galactic nuclei, are the primary mechanism by which energy, gas, and 
metals are injected into the IGM  \cite[e.g.,][]{veilleux05}. 

Observational information of galactic interaction and winds
comes in part from the observation of emission lines of hydrogen and metals 
\cite[e.g.,][]{bushouse87,veilleux05,strickland07}. While, 
these observations yield crucial information, such as the multiphase nature 
(from very hot ionized gas to cold neutral gas) and large-scale
morphology of these features, the detailed physics remain largely unknown. 
Because of their density-squared dependence, emission line observations 
are heavily biased toward the highest-density regions, which may only trace a 
relatively small fraction of the total mass and energy \cite[e.g.,][]{veilleux05,strickland00}. 
Measurements that rely on absorption lines are less biased to the highest
densities, but they require strong background sources.  Absorption line observations
of outflows beyond the Clouds probe only directions toward the 
brightest stellar clusters, losing detail as the stars are integrated
in the spectrograph slit. The Magellanic System is  near enough
to allow absorption line measurements to individual background objects. 

The Large Magellanic Cloud (LMC), Small Magellanic Cloud (SMC), and Milky Way (MW)
have long been suspected to have influenced each other. 
Large-scale mapping in \hi\ 21-cm emission of these clouds have revealed
several large gaseous structures, signatures of such interactions \cite{putman98,bruns05}.  
From the \hi\ maps of the Magellanic System, many other  distinct \hi\ features 
are visible: the Magellanic Bridge linking the SMC
and LMC, the Magellanic Stream, a 10\degr$\times$100\degr\ \hi\ filament
that trails the Clouds,  and the Leading Arm
that leads the Clouds in their orbit \cite{putman98}.  Thus the Magellanic System is 
an excellent laboratory for studying outflow (see below), accretion, tidal effects, and coronal gas
in and around low mass, low metallicity galaxies.

Over the last 8 years, {\fuse}\ has collected over 230 spectra of 
early-type stars in the Small and Large Magellanic Clouds (SMC and LMC) as 
well as a handful of QSOs behind the Stream and Bridge, providing a  high-quality
FUV LMC-SMC database. This database is extremely important because the multiphase 
nature of these features. The FUV  bandpass provides diagnostics of a wide range 
of gas phases, from the molecular clouds, to the neutral atoms, to low-, intermediate-,
and highly-ionized gas. In particular the importance of the ionized component 
has been truly revealed by {\fuse}. Although the spectral resolution of \fuse\ 
($R\sim 15,000$, $\delta v \simeq 20$ \km) is not so great for interstellar studies, it is high enough to 
estimate the column density and kinematics in many of these features and to decipher 
blueshifted or redshifted absorption relative to systemic velocities of the SMC and LMC (i.e., to find
signatures of outflow or accretion). 

Below I briefly touch on some of the \fuse\ successes. Because of space I do not discuss the
details (but I invite the reader to check out the various papers mentioned here) 
or other important results on the SMC and LMC  interstellar disk or stellar work 
that \fuse\ provided over the years \cite[e.g.,][]{massa03,walborn02,crowther02,tumlinson02,mallouris01}.

\section{Coronal Gas in the LMC, SMC, and Beyond}\label{sec-hot}
Prior to \fuse, little was known about the existence, not to mention 
the characteristics and distribution, of the coronal gas about the LMC.
Early {\it IUE}\ observations presented by \cite{savage79} were not conclusive because
the low data quality and owing to possible stellar contamination. First non-controversial
evidence was through the detection of \civ\ toward a couple of B-type stars
\cite{wakker98}. However, it is with the advent of \fuse\ that the  
hot halo about the LMC was established using the \ovi\ coronal diagnostic \cite{howk02}. 
Further evidence of outflow feeding the coronal gas of the LMC was recently presented by 
\cite{lehner07} who found a systematic similarity in the \ovi/\civ\ ratio
between the LMC and blueshifted high-velocity components at $100$--150 \km\ (incompatible with 
those of the LMC disk) seen in the UV absorption, suggesting that the blueshifted component 
has its origins in the LMC. 
Since the velocities of the blueshifted component relative to the LMC disk are larger than the
escape velocity of the LMC, this material may be escaping the LMC, polluting the intergalactic 
space between the LMC and the Milky Way or serving as fuel for the Magellanic Stream \cite{nidever08}.
Independently, \cite{staveley-smith03} also found that the relatively high \hi\ column density
clouds at $v_{\rm LSR} \sim \,$100--160 \km\ 
are often seen projected onto \hi\ voids in the LMC disk and are connected 
to the disk with spatial and kinematic bridges in position-velocity plots, 
suggesting outflows as well (see also below). 

In the SMC, early {\em IUE}\ observations allowed to provide some evidence for a global component
of highly ionized gas in the SMC \cite{fitz85}, but \cite{hoopes02} truly established the presence
of a substantial and extended component of \ovi\ coronal gas in the SMC.  They also found
that the large star-forming regions in the SMC strongly affect the distribution
of hot gas, and more so than in the LMC \cite[][but see also Blair et al., these proceedings]{howk02,lehner07}. 
Possible origins for the coronal gas include superbubbles and galactic fountains. 

Finally, \cite{sembach03}, and see also \cite{fox06}, show that some high-velocity \ovi\ 
clouds were associated with the Magellanic Stream. This indicates that the stream 
extends further out in space than the regions sampled by \hi\ 21 cm emission, suggesting
coronal gas up to 50 kpc \cite[see also][]{collins05,fox05}.  Hence while it was
often assumed before the {\fuse}\ results, it is now established that these galaxies are
surrounded by hot coronal gas on large scale. 

\section{The Magellanic Stream and Leading Arm}
Unfortunately, there are not many QSOs or AGNs behind the Stream, and most
of our knowledge in the UV came from GHRS and STIS observations \cite[e.g.][]{gibson00}. Nevertheless 
one notable exception was the \fuse\ observation of NGC\,3783, which probes gas in the
leading Arm of the Stream \cite{sembach01}. Thanks to a large number of spectral diagnostics, these authors show 
that ionization corrections were small and the abundance
pattern suggests that the Leading Arm contains dust grains that have been processed
significantly.  The most important finding was the discovery of 
H$_2$ associated with this high-velocity cloud (HVC). This is the only HVC where
molecules have been found so far, further supporting that the gas from the stream was tidally 
pulled or ejected from a galaxy based on  H$_2$ formation time-scale \cite{sembach01}. 

\section{The Magellanic Bridge}
Although only two lines of sight could be observed with \fuse, our knowledge 
of the conditions in the Bridge has expanded thanks to these observations. In a series 
of papers, my collaborators and I show that the diffuse  gas in the Bridge is multiphase, 
consisting of neutral, weakly, and highly ionized gas using a combination of STIS
and \fuse\ data \cite{lehner01b,lehner02,lehner08}. 
\fuse\ observations toward one star embedded in the Bridge show that a small amount of molecular 
hydrogen exists in this very much ionized environment
(see below). Detection of CO molecules were also subsequently found \cite{muller03}, 
but in a much denser region of the Bridge, a.k.a. the SMC wing \cite[see ][]{lehner08}. 

\fuse\ also provided several \oi\ lines that were crucial for determining the
first gas-phase metallicity of the Magellanic Bridge (combined with STIS),
$[{\rm Z/H}] = -1.02 \pm 0.07$ toward one sightline, and $-1.7 < [{\rm Z/H}] < -0.9$
toward another one \cite{lehner08}. These are in excellent agreement with B-type 
stellar abundances in the Bridge \cite{rolleston99,lee05}. If we believe current
tidal models, then the Bridge is only 200 Myr old and was pulled from the
SMC ($[{\rm Z/H}]_{\rm SMC} = -0.6$), which implies that
the diffuse gas was highly diluted with extremely low metallicity gas. 
On the other hand the very low present-day metallicity in the Bridge
is similar to the SMC before its burst of star formation that occurred
about 2.5 Gyr ago and a time that coincidentally corresponds as well to a close encounter 
between the SMC and LMC.  This may not only be a pure coincidence since
interaction between galaxies create bursts of star
formation within the interacting galaxies. Although some chemico-dynamical 
models of the LMC-SMC-Galaxy interactions attempt to address that 
question \cite[e.g.,][]{bekki07}, I do not think there is yet a satisfactory model
that explains the stellar and interstellar results. 
Another issue is that current models only attempt to model the neutral
component, while we find that 80\% of the gas may be ionized, 
implying that the largest fraction of the gas mass 
of the Bridge may come from the ionized gas \cite{lehner08}. 

\section{HVCs in Front the Magellanic Clouds}
Finally, I will conclude on the HVC complex observed between the MW and the Clouds at LSR 
velocities between about 90 and 150 \km, already mentioned in \S\ref{sec-hot}. 
Paradoxically, even though these HVCs have many background stars ($>230$) that can be used to 
provide a detailed study of its abundances, distribution, etc., 
little is still known about this complex. My collaborators  (Chris Howk 
and Lister Staveley-Smith) and I are working on a series of papers
that should shed light on these HVCs. 
Toward the Bridge, we first reported the existence of an HVC that is fully ionized
using STIS observations \cite{lehner01a}, 
at a time when HVCs were mostly known as neutral entities. When {\fuse}\ came online
it was realized that these HVCs were observed frequently in the spectra of LMC and 
SMC stars in the low and high ions \cite{danforth02,hoopes02,howk02}. 

Toward the LMC, independent results show that this HVC may be linked to outflows from the LMC
(see \S\ref{sec-hot}). One way to further test this is to estimate the metallicities
of these HVCs. Combining \hi\ emission and \oi\ absorption data toward about 80 sightlines, 
our preliminary results show that  about 88\% of our sample has an O abundance relative to solar between  $-0.8$ 
and $-0.2$ dex, and peaking near $-0.5$ dex. 
The average  metallicity  of $-0.5$ dex relative to solar abundance seems to support
the recent claims of \cite{nidever08} that massive outflow had to occur
over 1.3 Gyr ago, since chemical evolution models of LMC predict it had a metallicity
of  $-0.5$ dex approximately 2 Gyr ago. The scatter 
in the metallicities is also quite similar to those found in the Stream \cite{gibson00}, 
and suggests that outflows are still ongoing and that part of the gas 
may also be mixed with a lower metallicity component. In the future, we will fully characterize
the properties of this HVC complex. 

\section{Concluding Thoughts}
In summary, it has been a nice ride with many exciting results that will continue 
to grow thanks to the rich SMC/LMC \fuse\ archive. 
 Future UV observations with COS will provide 
new information, e.g. we will be able to probe the characteristics of
the LMC/SMC halo throughout their coronal gas at much larger impact parameters. But I believe
the \fuse\ results already provide key information that models of the Magellanic Cloud-Milky Way system
should attempt to reproduce and use. 

Since the theme of this conference is also the future of UV 
and  since the UV is at the core of astrophysical inference with uniquely rich spectral
diagnostics, we need to think possibly not so much in terms of better resolution (although that is
always good!), but to bigger and more sensitive instruments (even than COS) to go beyond the Clouds. 
As accretion and outflow are so important for our understanding  
galaxies and input physics in cosmological simulations, an unbiased sample of galaxies in
the Local Group where these processes may not be as entangled as in the LMC/SMC and where UV spectroscopy 
can be achieved toward individual background objects will be key for making more progress 
in our understanding of the influence of accretion/outflow in galaxy evolution and in modeling these processes. 
It is also important to realize that many of the \fuse\ observations relied on STIS E140M data
(e.g., Ly$\alpha$, \nv, \civ, etc...), so a future UV space mission will need somehow to combine 
the FUV and UV. 

As a final word, I want to thank you the \fuse\ 
scientific, planning, and engineering  teams for their dedicated and creative efforts that allow this 
telescope to exist and ran  well beyond its nominal 3-year mission.


\begin{thebibliography}{9}

\bibitem{bekki07} 
Bekki, K., \& Chiba, M.\ 2007, PASA, 24, 21 

\bibitem{bruns05} 
Br{\"u}ns, C., et al.\ 2005, A\&A, 432, 45 

\bibitem{bushouse87} 
Bushouse, H.~A.\ 1987, ApJ, 320, 49 

\bibitem{collins05} Collins, J.~A., Shull, 
J.~M., \& Giroux, M.~L.\ 2005, ApJ, 623, 196 

\bibitem{crowther02} Crowther, P.~A., 
Hillier, D.~J., Evans, C.~J., Fullerton, A.~W., De Marco, O., 
\& Willis, A.~J.\ 2002, ApJ, 579, 774 

\bibitem{danforth02} Danforth, C.~W., Howk, 
J.~C., Fullerton, A.~W., Blair, W.~P., 
\& Sembach, K.~R.\ 2002, ApJs, 139, 81 

\bibitem{fitz85} 
Fitzpatrick, E.~L., \& Savage, B.~D.\ 1985, ApJ, 292, 122 

\bibitem{fox06} Fox, A.~J., Savage, B.~D., \& Wakker, B.~P.\ 2006, ApJS, 165, 229 

\bibitem{fox05} Fox, A.~J., Wakker, B.~P., 
Savage, B.~D., Tripp, T.~M., Sembach, K.~R., 
\& Bland-Hawthorn, J.\ 2005, ApJ, 630, 332 

\bibitem{gibson00} Gibson, B.~K., Giroux, 
M.~L., Penton, S.~V., Putman, M.~E., Stocke, J.~T., 
\& Shull, J.~M.\ 2000, AJ, 120, 1830 

\bibitem{hoopes02} Hoopes, C.~G., Sembach, 
K.~R., Howk, J.~C., Savage, B.~D., 
\& Fullerton, A.~W.\ 2002, ApJ, 569, 233 

\bibitem{howk02} Howk, J.~C., Sembach, 
K.~R., Savage, B.~D., Massa, D., Friedman, S.~D., 
\& Fullerton, A.~W.\ 2002, ApJ, 569, 214 

\bibitem{larson78} Larson, R.~B., \& Tinsley, B.~M.\ 1978, ApJ, 219, 46 

\bibitem{lee05} 
Lee, J.-K., Rolleston, W.~R.~J., Dufton, P.~L., \& Ryans, R.~S.~I.\ 2005, A\&A, 429, 1025 

\bibitem{lehner02} 
Lehner, N.\ 2002, ApJ, 578, 126 

\bibitem{lehner08} 
Lehner, N., Howk, J.~C., Keenan, F.~P., \& Smoker, J.~V.\ 2008, ApJ, 678, 219 

\bibitem{lehner07} 
Lehner, N., \& Howk, J.~C.\ 2007, MNRAS, 377, 687 

\bibitem{lehner01a} 
Lehner, N., Keenan, F.~P., \& Sembach, K.~R.\ 2001a, MNRAS, 323, 904 

\bibitem{lehner01b} 
Lehner, N., Sembach, K.~R., Dufton, P.~L., Rolleston, W.~R.~J., 
\& Keenan, F.~P.\ 2001b, ApJ, 551, 781 

\bibitem{mallouris01} Mallouris, C., et 
al.\ 2001, ApJ, 558, 133 

\bibitem{massa03} Massa, D., Fullerton, 
A.~W., Sonneborn, G., \& Hutchings, J.~B.\ 2003, ApJ, 586, 996 

\bibitem{muller03} 
Muller, E., Staveley-Smith, L., \& Zealey, W.~J.\ 2003, MNRAS, 338, 609 

\bibitem{nidever08} Nidever, D.~L., 
Majewski, S.~R., \& Burton, W.~B.\ 2008, ApJ, 679, 432 

\bibitem{putman98} Putman, M.~E., et al.\ 
1998, Nature, 394, 752 

\bibitem{rolleston99} Rolleston, W.~R.~J., Dufton, P.~L., McErlean, N.~D., \& Venn, K.~A.\ 1999, A\&A, 348, 728 

\bibitem{savage79} Savage, B.~D., \& de Boer, K.~S.\ 1979, ApJL, 230, L77 

\bibitem{simcoe06} Simcoe, R.~A., Sargent, 
W.~L.~W., Rauch, M., \& Becker, G.\ 2006, ApJ, 637, 648 

\bibitem{staveley-smith03} 
Staveley-Smith, 
L.,  Kim, S., Calabretta, M.~R., Haynes, R.~F., 
\& Kesteven, M.~J.\ 2003, MNRAS, 339, 87 

\bibitem{sembach01} 
Sembach, K.~R., Howk, 
J.~C., Savage, B.~D., \& Shull, J.~M.\ 2001, AJ, 121, 992 

\bibitem{sembach03} Sembach, K.~R., et al.\ 
2003, ApJS, 146, 165 

\bibitem{strickland07} Strickland, D.~K., \& Heckman, T.~M.\ 2007, ApJ, 658, 258 

\bibitem{strickland00} Strickland, D.~K., \& Stevens, I.~R.\ 2000, MNRAS, 314, 511 

\bibitem{tumlinson02} Tumlinson, J., et 
al.\ 2002, ApJ, 566, 857 

\bibitem{veilleux05} Veilleux, S., Cecil, G., \& Bland-Hawthorn, J.\ 2005, ARA\&A, 43, 769 

\bibitem{wakker98} Wakker, B., Howk, J.~C., 
Chu, Y.-H., Bomans, D., \& Points, S.~D.\ 1998, ApJ, 499, L87 


\bibitem{walborn02} Walborn, N.~R., 
Fullerton, A.~W., et al. \ 2002, ApJS, 141, 443 







\end{thebibliography}
\end{document}